\documentclass[%
 reprint,
superscriptaddress,
 amsmath,amssymb,
 aps,
prb,
longbibliography
]{revtex4-2}

\usepackage{graphicx}%
\usepackage{dcolumn}%
\usepackage{multirow}
\usepackage{amssymb}
\usepackage{comment}
\usepackage{MnSymbol,wasysym}
\usepackage{setspace}
\usepackage{braket}

\newcommand{\LC}[1]{\textcolor{blue}{#1}}

\usepackage{soul}

\usepackage[dvipsnames,table,xcdraw]{xcolor}

\usepackage{hyperref}
\hypersetup{
    colorlinks=true,
    linkcolor=Blue,
    filecolor=Blue,      
    urlcolor=Blue,
    citecolor=blue,
}

\renewcommand\vec[1]{\ensuremath\boldsymbol{#1}} 

\begin{abstract}
Charge-density wave (CDW) orders are conventionally described as modulations of on-site charge at an ordering wave vector $\vec{Q}$ with symmetry-related wave vectors typically forming a multicomponent order-parameter manifold.
Recent developments significantly broadened this phenomenology to bond and loop-current density waves, which possess nontrivial, potentially symmetry-breaking textures within the unit cell in addition to their spatial modulation at $\vec{Q}$.
Such textures arise from particle-hole condensates with nonzero angular momentum, analogous to unconventional superconductivity, and their symmetries are described by the little group $G_{\vec{Q}}$.
In this work, we develop a framework for unconventional CDW phases that simultaneously incorporates the local symmetries described by little group and the presence of multiple symmetry-related ordering wave vectors, also known as the star.
The resulting multicomponent order parameter transforms under representations of the full space group induced from irreducible representations of $G_{\vec{Q}}$.
We apply this framework to two-dimensional lattices with sixfold symmetry and ordering wave vectors along high-symmetry lines, and derive the corresponding Landau free energies.
As a microscopic example, we demonstrate the emergence of unconventional bond and loop-current orders from electronic interactions on the triangular lattice within the random phase approximation, and determine their ground states by microscopically evaluating the relevant coefficients in the free energy.
Our framework provides a systematic route to describing unconventional modulated phases and can be readily extended to more complex lattices.
\end{abstract}

\begin{document}
\title{Unconventional bond- and current-density waves on hexagonal lattices}

\author{Andr\'as L. Szab\'o}
\affiliation{Max Planck Institute for Solid State Research, D-70569 Stuttgart, Germany}
\affiliation{Institute for Theoretical Physics, ETH Zurich, 8093 Zurich, Switzerland}

\author{Hannes Braun}
\affiliation{Max Planck Institute for Solid State Research, D-70569 Stuttgart, Germany}
\affiliation{School of Natural Sciences, Technische Universit\"at München, D-85748 Garching, Germany}

\author{Laura Classen}
\affiliation{Max Planck Institute for Solid State Research, D-70569 Stuttgart, Germany}
\affiliation{School of Natural Sciences, Technische Universit\"at München, D-85748 Garching, Germany}

\date{\today}

\maketitle

\section{Introduction}

Charge-density wave (CDW) orders arise in many correlated quantum materials from electronic or phononic instabilities.
Their study provides insight into the fundamental mechanisms underlying the coupling between electrons and lattice in materials.
CDWs describe states of broken translation symmetry with a modulation wave vector $\vec{Q}$.
In the simplest case, this modulation applies to the on-site charge density, whereas in more complex scenarios the CDW manifests as the spatial modulation of bonds, orbitals, or loop currents around lattice sites.
Exotic charge order has been discussed in systems such as kagome metals~\cite{Jiang2021,Kang2022,Xing2024,Candelora2026, Denner2021, Christensen2021}, transition metal dichalcogenides such as TiSe$_2$~\cite{Ishioka2010,vanWezel2011,HKim2024,Kim2024,Jiang2026,Edwards2026}, or rare-earth tritellurides~\cite{Wang2022,Singh2025,Wulferding2025}.

The aforementioned more complex CDWs feature a nontrivial texture of bonds or currents on the atomic scale, which is additionally modulated on a wavelength associated with $\vec{Q}$. 
Microscopically, such states arise from a particle-hole condensate with angular momentum $l>0$. Thus, they can be thought of as 
\emph{unconventional} charge density waves (UCDWs) \cite{PhysRevB.62.4880,Braun2025}, in analogy with their superconducting counterparts.
In addition to their locally nontrivial characteristics, similarly to conventional density wave orders, UCDWs generally also feature multiple symmetry related modulation vectors $\vec{Q}_i$, spanning a multicomponent order-parameter space.
This makes such phases particularly interesting, as it can further accommodate a complex landscape of ordered states.
On the one hand, the CDW ground state can break further lattice symmetries via selecting a particular set of $\vec{Q}_i$, such as proposed for $A$V$_3$Sb$_5$ kagome metals ($A=$ K, Rb, Cs)~\cite{Kang2022,Christensen2021}.
Density waves can also facilitate the emergence of more exotic states of matter, such as intertwined and vestigial orders~\cite{Kivelson1998,Fradkin2015,Fernandes2019}.
In this context, superconductivity has been known to emerge on a background of exotic CDWs producing intertwined pair-density wave phases in materials platforms such as $2H$-NbSe$_2$~\cite{Liu2021} or CsV$_3$Sb$_5$~\cite{Chen2021}.

While this multicomponent nature is well described by the set of symmetry-related momenta, also called the \emph{star} of $\vec{Q}$, the local features are in turn described by the little group (LG) $G_{\vec{Q}}$, which is a subgroup of the full space group $G$ and encompasses operations that leave $\vec{Q}$ invariant modulo reciprocal lattice vectors.
Note that the vast majority of commonly studied density wave orders occur at a point or along a line of symmetry in the Brillouin zone (BZ), implying a LG with at least one irreducible representation (irrep) transforming nontrivially in $G_{\vec{Q}}$.
Microscopically, such nontrivial local symmetries manifest exactly via textures on the scale of the unit cell, such as bonds, currents, or sublattice, and arise from form factors with $l>0$. 
Notice that while common experimental techniques such as scattering and surface-sensitive probes readily identify density-wave orders through the emergence of new Fourier peaks at the ordering vectors, they typically provide limited direct information about the LG irrep and the associated local symmetry of the ordered state.
Therefore, a description of density wave orders that incorporates the interplay between the LG and the star of $\vec{Q}$ constitutes an important step towards a comprehensive characterization of such phases.

\begin{figure}
    \centering
    \includegraphics[width=0.8\linewidth]{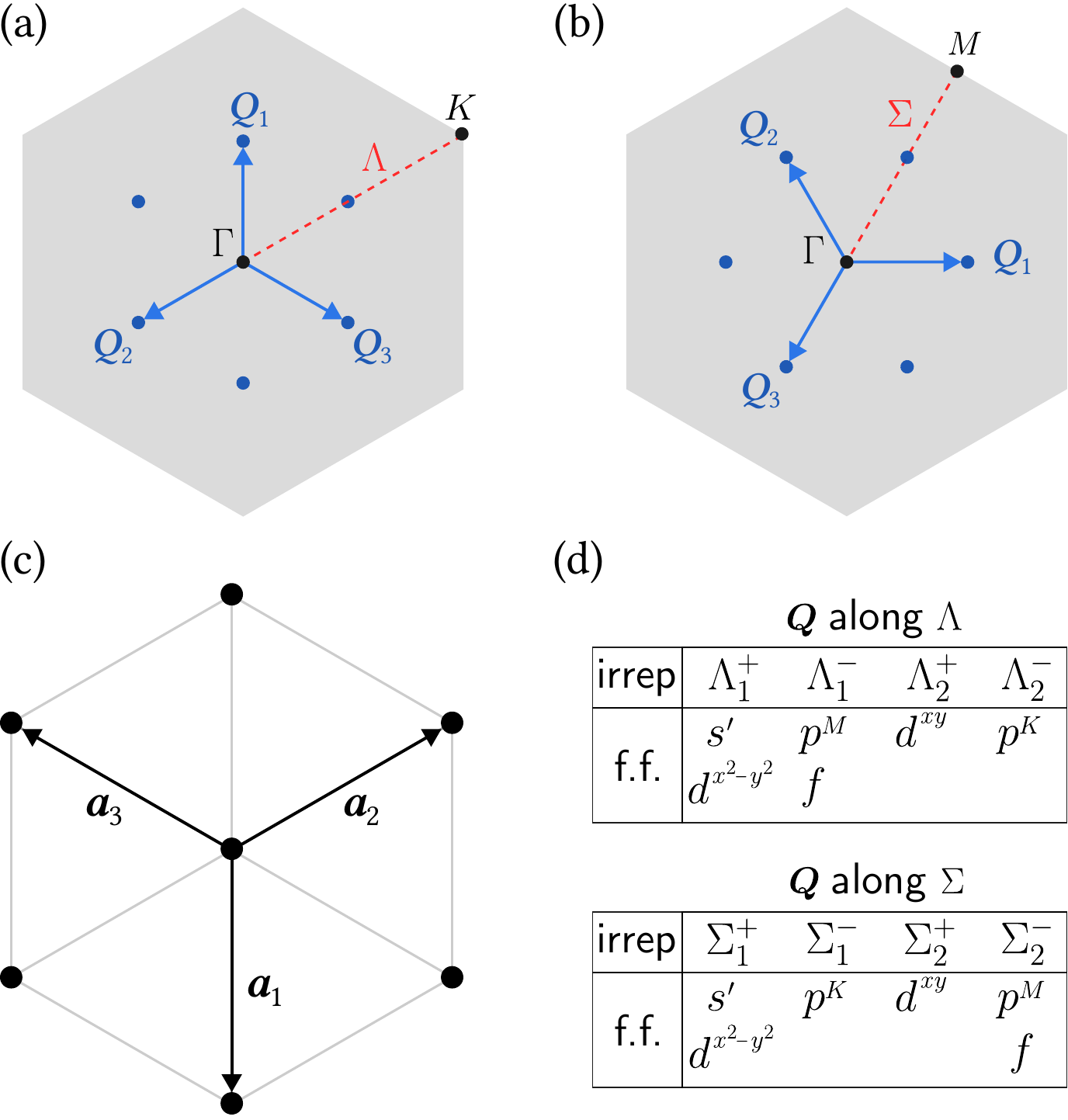}
    \caption{Top: hexagonal Brillouin zone and $\pm \vec{Q}_{1,2,3}$ symmetry related ordering wave vectors forming the star. In (a) these fall on the $\Gamma-K$ line also known as $\Lambda$ and in (b) on the $\Gamma-M$ or $\Sigma$ line. Bottom: (c) Triangular lattice with bonds $\vec{a}_{1,2,3}$ indicated with arrows. (d) Classification of nearest neighbor form factors on the triangular lattice according to LG irreps for the two cases shown in (a) and (b). For the expressions of the form factors see text. }
    \label{fig:Lattice}
\end{figure}

Here, we study a family of UCDW orders arising on two-dimensional hexagonal lattices, with $\vec{Q}_i$ along high-symmetry lines in the BZ.
Specifically, a set of six ordering vectors $\pm\vec{Q}_i$ with $i=1,2,3$ forming the star lie along the line connecting the $\Gamma - K$ or $\Gamma - M$ points, also known as $\Lambda$ or $\Sigma$ lines, respectively, see Fig.~\ref{fig:Lattice}.
The corresponding LG contains the identity and a mirror operation through the high-symmetry line, implying the existence of two local irreps, even and odd under this mirror, respectively.
The six-component order parameter transforms according to representations of the space group (SG) induced from either irrep of the LG, which we analyze on group-theoretical grounds.
In doing so, we consider both time-reversal symmetry (TRS) preserving bond and TRS-breaking current orders.
Studying the corresponding Landau free energy allows us to determine the circumstances favoring either a unidirectional or a multi-$Q$ ground state.
As a minimal example featuring a single sublattice, we consider the triangular lattice, and showcase the real-space pattern of various bond and current orders.
Nevertheless, our free energy is valid to a larger class of hexagonal lattices such as the honeycomb or kagome lattice, and our methodology is readily generalizable to more complex space groups.

In a recent work, it was argued using random phase approximation (RPA) and functional renormalization group that such density waves can indeed be stabilized on the triangular lattice via an electronic mechanism, when the chemical potential is tuned away from the van Hove filling~\cite{Braun2025}.
To connect with our symmetry analysis here, we identify the leading bond- and current-density-wave instabilities within RPA for Fermi energies below and above Van Hove filling, and determine the corresponding ground states by computing the relevant coefficients appearing in the Landau free energy.

\begin{figure*}
    \centering
    \includegraphics[width=0.9\linewidth]{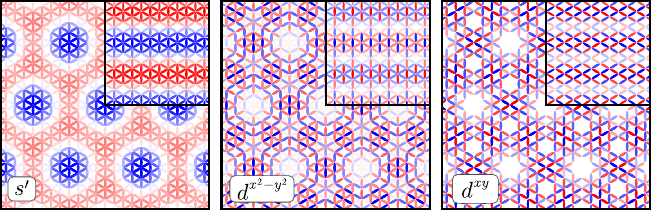}
    \caption{Depictions of bond orders in real space with form factors indicated in the bottom left. In each panel, the main figure and the inset correspond to a multi-$Q$ and single-$Q$ ground state, respectively, and we set all phases to be $\phi_i=\theta_i=0$. }
    \label{fig:bonds}
\end{figure*}

\section{Symmetry analysis}

\begin{figure*}
    \centering
    \includegraphics[width=0.9\linewidth]{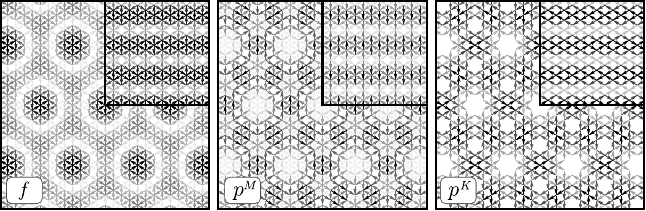}
    \caption{Depictions of current orders in real space with form factors indicated in the bottom left. In each panel, the main figure and the inset correspond to a multi-$Q$ and single-$Q$ ground state, respectively, and we set all phases to be $\phi'_i=\theta'_i=0$. }
    \label{fig:currents}
\end{figure*}

As starting point, we consider the mean-field Hamiltonian describing the ordered phase in the general form
\begin{align}
    \mathcal{H}_{\rm CDW} = \sum_{\vec{k},\sigma} \sum_{i=1,2,3} \Big( &\Delta_{+i} F_{+i}(\vec{k}+\vec{Q}_i/2) c^\dag_{\vec{k}+\vec{Q}_i,\sigma}c_{\vec{k},\sigma} \label{eq:HMF} \\
    + &\Delta_{-i} F_{-i}(\vec{k}-\vec{Q}_i/2) c^\dag_{\vec{k}-\vec{Q}_i,\sigma}c_{\vec{k},\sigma} + {\rm H.c.}\Big),\nonumber
\end{align}
where $c_{\vec{k},\sigma}$ annihilates an electron with momentum $\vec{k}$ and spin $\sigma$ in a given band, and $\pm \vec{Q}_i$ lie along the $\Lambda$ or $\Sigma$ line, see Fig.~\ref{fig:Lattice}.
The momentum-dependent functions $F_{\pm i}(\vec{k})$ are local form factors, which we evaluate at the average of the incoming and outgoing momenta, as this convention makes the transformation properties under TRS more transparent.
In what follows, the six complex components $\boldsymbol{\Delta}=(\Delta_{\pm 1}, \Delta_{\pm 2}, \Delta_{\pm 3})$ constitute our order parameters in the Landau formalism.
To build a phenomenological theory, we first analyze their transformation properties under space-group symmetries of the lattice, alongside TRS. For concreteness, we consider the triangular lattice, which allows us to specify form factors and depict the different bond and current orders in real space (Figs.~\ref{fig:bonds},\ref{fig:currents}).
We note, however, that the symmetry classification via the LG irreps is valid more generally for two-dimensional systems with a hexagonal symmetry after an appropriate redefinition of order parameters and form factors.

\subsection{Space-group symmetry}\label{sec:space_group_symmetry}

We consider two-dimensional hexagonal lattices with $C_{6v}$ symmetry.
The point group (PG) can be generated by a sixfold rotation $C_6$, and a diagonal mirror $M_x$, acting on the coordinate as $x\to -x$.
It contains all together six conjugacy classes: $E$ (identity), $2C_6$, $2C_3$ and $C_2$ (three- and twofold rotations), as well as $3\sigma_d$ and $3\sigma_v$ (diagonal and vertical mirrors), with the number indicating the size of each class.
As density-wave orders break translational symmetry, PG operations must be augmented by primitive lattice vector translations of the corresponding lattice, described by an appropriate space group.
To obtain the relevant representation theory, we embed the two-dimensional symmetry group into the three-dimensional space group $P6mm$, whose point group is isomorphic to $C_{6v}$ and restrict to the $k_z=0$ plane.
Here, the $\Lambda$ and $\Sigma$ lines respectively connect the $\Gamma-K$ and $\Gamma-M$ points of the BZ (excluding both end points).
All momenta along the high-symmetry lines have the same symmetry, with the caveat that rational and irrational fractions of the momentum at $K$ or $M$ yield respectively commensurate and incommensurate orders.
Throughout we keep the discussion general, and comment on commensurability where appropriate.

For the sake of concreteness, we consider the symmetries of the $\Lambda$ line, which connects with our microscopic example (see Section.~\ref{sec:microscopic_model}), where the instability is found near $\vec{Q}=\vec{K}/4$, where $\vec{K}$ is the momentum at $K$.
The theory along the $\Sigma$ line can be obtained via straightforward redefinitions, as we comment on along the way.
Let us pick $\vec{Q}_1 \parallel k_y$ as the representative ordering wave vector, see Fig.~\ref{fig:Lattice}(a).
Then, the LG which leaves $\vec{Q}_1$ invariant consists of $\{E,M_x\}$.
It therefore supports two irreps $\Lambda_{1,2}$, respectively even and odd under $M_x$.
On the other hand, applying symmetry operations not contained in the LG (such as $C_6$ rotations) generates the star $\{\vec{Q}_{\pm 1},\vec{Q}_{\pm 2},\vec{Q}_{\pm 3} \}$.
More generally, the LG at symmetry-related momenta can be written as $\{E, \sigma_d \}$, where $\sigma_d$ is the conjugacy class containing mirrors through rotated diagonal planes.

The form factors $F_{\pm i}(\vec{k})$ in Eq.~(\ref{eq:HMF}) transform under the LG irreps at the corresponding momenta $\pm \vec{Q}_i$.
Specifically, for $\Lambda_1$, we consider $F_{\pm i}(\vec{k})$ to be one of  
the following nearest-neighbor form factors on the triangular lattice in Fig.~\ref{fig:Lattice} 
\begin{align}
   &s'_{\pm i}(\vec{k}) =\sqrt{\frac{2}{3}} \Big( \cos(\vec{k}.\vec{a}_i)+\cos(\vec{k}.\vec{a}_j)+\cos(\vec{k}.\vec{a}_k) \Big), \nonumber \\
   &p^M_{\pm i}(\vec{k})=\pm\frac{ 1}{\sqrt{3}} \Big( -2\sin(\vec{k}.\vec{a}_i)+\sin(\vec{k}.\vec{a}_j)+\sin(\vec{k}.\vec{a}_k) \Big), \nonumber \\
   &d^{x^2-y^2}_{\pm i}(\vec{k})=\frac{1}{\sqrt{3}} \Big( -2\cos(\vec{k}.\vec{a}_i)+\cos(\vec{k}.\vec{a}_j)+\cos(\vec{k}.\vec{a}_k) \Big), \nonumber \\
   &f_{\pm i} (\vec{k}) =\pm\sqrt{\frac{2}{3}} \Big( \sin(\vec{k}.\vec{a}_i)+\sin(\vec{k}.\vec{a}_j)+\sin(\vec{k}.\vec{a}_k) \Big), \label{eq:ff1}
\end{align}
where $i,j,k$ are cyclic permutations of $(1,2,3)$. 
In addition, the conventional momentum-independent $s$-wave form factor also transforms under $\Lambda_1$. 
The extended $s$-wave ($s'$), and $d^{x^2-y^2}$-wave functions are even parity, whereas $p^M$ ($p$-wave with nodes along the $M$ points) and $f$-wave form factors are odd parity.
For $\Lambda_2$, we consider
\begin{align}
   p^K_{\pm i}(\vec{k})&=\pm\Big( -\sin(\vec{k}.\vec{a}_j)+\sin(\vec{k}.\vec{a}_k) \Big), \nonumber \\
   d^{xy}_{\pm i}(\vec{k})&= -\cos(\vec{k}.\vec{a}_j)+\cos(\vec{k}.\vec{a}_k) , \label{eq:ff2}
\end{align}
with the $p^K$ ($p$-wave with node along the $K$ points) and $d^{xy}$-waves being odd and even under inversion, respectively.
Notice that inversion $\vec{k}\to -\vec{k}$ is not part of the LG, therefore different parities can mix under $\Lambda_{1,2}$.
Nevertheless, TRS differentiates between inversion even and odd form factors, as we highlight later.

While a single component $\Delta_{i}$ transforms according to an irrep $\Lambda_1$ or $\Lambda_2$ of the LG at some $\vec{Q}_i$, the symmetry of the modulated phase is determined by the transformation properties of the six-component order parameter $\boldsymbol{\Delta}$ under the full SG, which are described by the representations of the SG induced from $\Lambda_{1,2}$.
An induced representation naturally extends an irrep of the LG to the full SG by incorporating all wave vectors in the star of $\vec{Q}_i$~\cite{Bradley2009}.
In practice, we infer the transformation properties by applying every element of the SG on Eq.~(\ref{eq:HMF}), regardless of whether it belongs to the LG, and subsequently absorbing the resulting transformation into the components $\Delta_{i}$.

Since $P6mm$ is symmorphic, we can assess translations and PG operations separately.
We apply a PG-type operation $g$ on Eq.~(\ref{eq:HMF}) as
\begin{align}
&\sum_{k,\sigma} \Delta_{\pm i} F_{\pm i}[(\vec{k}\pm \vec{Q}_i/2)\vec{a}] c^\dag_{g\vec{k}\pm g\vec{Q}_i,\sigma}c_{g\vec{k},\sigma} \nonumber \\
=&\sum_{p,\sigma}  \Delta_{\pm i} F_{\pm i}[(\vec{p}\pm g\vec{Q}_i/2)g\vec{a}] c^\dag_{\vec{p}\pm g\vec{Q}_i,\sigma}c_{\vec{p},\sigma}
\end{align}
where $\vec{a}\in \{\pm \vec{a}_{1,2,3} \}$.
Above we introduced $\vec{p}=g\vec{k}$ as new summation variable, such that the argument of $F_{\pm i}$ becomes $(g^{-1}\vec{p}\pm g^{-1} g\vec{Q}_i/2)\vec{a}$, and used the fact that $g$ is represented by an orthogonal matrix.
Therefore, $g$ can be implemented by applying it to the $\vec{Q}_i$ ordering wave vectors, as well as the bonds appearing in the form factors.
Note that translations by a real-space vector $\vec{t}$ act as  $T_{\vec{t}}c_{\vec{k},\sigma}T_{\vec{t}}^{-1}=e^{-i\vec{kt}}c_{\vec{k},\sigma}$, such that the order parameter transforms as $\Delta_{\pm i} \to e^{\pm i \vec{Q}_i.\vec{t} } \Delta_{\pm i}$, independent of the form factor.

In the following we adapt our notation to distinguish between the two irreps $\Lambda_1$ and $\Lambda_2$, and  introduce $\Delta_{\pm i}\equiv \psi_{\pm i}$ or $\Delta_{\pm i}\equiv\eta_{\pm i}$  whenever $F_{\pm i}$ is in $\Lambda_1$ and $\Lambda_2$, respectively.
The transformation properties of the above order parameters under $C_6$, $M_x$, and $M_y$, alongside the characters of the induced representation are shown in Table~\ref{tab:C6vtransformations}.
We find that $\psi_{\pm i}$ transform the same way as the coordinate or momentum, which is consistent with $\Lambda_1$ containing the trivial $s$-wave form factor.
On the other hand, $\eta_{\pm i}$ acquire an additional minus sign under both vertical and diagonal mirrors.
\begin{table}[h]
\centering
\renewcommand{\arraystretch}{1.1}
\setlength{\tabcolsep}{3pt}
\begin{tabular}[t]{|c|ccc|}
\hline
$\Lambda_1$ & $C_6$ & $M_x$ & $M_y$ \\
\hline
$\psi_{+1}$ & $\psi_{-3}$ & $\psi_{+1}$ & $\psi_{-1}$ \\
$\psi_{-1}$ & $\psi_{+3}$ & $\psi_{-1}$ & $\psi_{+1}$ \\
$\psi_{+2}$ & $\psi_{-1}$ & $\psi_{+3}$ & $\psi_{-3}$ \\
$\psi_{-2}$ & $\psi_{+1}$ & $\psi_{-3}$ & $\psi_{+3}$ \\
$\psi_{+3}$ & $\psi_{-2}$ & $\psi_{+2}$ & $\psi_{-2}$ \\
$\psi_{-3}$ & $\psi_{+2}$ & $\psi_{-2}$ & $\psi_{+2}$ \\
\hline
\end{tabular}
\hspace{0.1cm}
\setlength{\tabcolsep}{3pt}
\begin{tabular}[t]{|c|ccc|}
\hline
$\Lambda_2$  & $C_6$ & $M_x$ & $M_y$ \\
\hline
$\eta_{+1}$ & $\eta_{-3}$ & $-\eta_{+1}$ & $-\eta_{-1}$ \\
$\eta_{-1}$ & $\eta_{+3}$ & $-\eta_{-1}$ & $-\eta_{+1}$ \\
$\eta_{+2}$ & $\eta_{-1}$ & $-\eta_{+3}$ & $-\eta_{-3}$ \\
$\eta_{-2}$ & $\eta_{+1}$ & $-\eta_{-3}$ & $-\eta_{+3}$ \\
$\eta_{+3}$ & $\eta_{-2}$ & $-\eta_{+2}$ & $-\eta_{-2}$ \\
$\eta_{-3}$ & $\eta_{+2}$ & $-\eta_{-2}$ & $-\eta_{+2}$ \\
\hline
\end{tabular}
\vspace{0.8cm}

\centering
\begin{tabular}{|c|cccccc|}
\hline
	irrep & $E$ & $2 C_6$ & $2 C_3$ & $C_2$ & $3 \sigma_d$	& $3 \sigma_v$ \\
    \hline
$\Lambda_1$ & 6 & 0 & 0 & 0 & $+2$ & 0 \\
$\Lambda_2$ & 6 & 0 & 0 & 0 & $-2$ & 0 \\
\hline
\end{tabular}

\caption{Top: transformation properties of the order parameter components $\psi_{\pm i}$ and $\eta_{\pm i}$ under representative elements of the PG $C_{6v}$. Bottom: characters of the representations of SG $P6mm$ induced from the irreps $\Lambda_{1,2}$ of the LG at $\vec{Q}_1$.
We obtain the induced representations via extending the LG irrep to the full SG, see text.}
\label{tab:C6vtransformations}
\end{table}

\subsection{Time-reversal symmetry}

We now move on to incorporate TRS, which acts as $\mathcal{T} h(\vec{k}) c^{(\dagger)}_{\vec{k},\sigma} \mathcal{T}^{-1}= h^*(\vec{k}) c^{(\dagger)}_{-\vec{k},\bar{\sigma}}$ for some prefactor $h(\vec{k})$ which may depend on momentum.
We find that even and odd-parity form factors are even and odd under TRS, respectively.
Then, each irrep $\Lambda_{1,2}$ further splits into time-reversal even $\Lambda_{1,2}^+$ and odd $\Lambda_{1,2}^-$, and the corresponding order parameters describe bond and current orders, respectively.
The form factors accordingly split based on their parity, such that trivial and extended $s$-wave, as well as $d^{x^2-y^2}$ transform under $\Lambda_1^+$, whereas $p^M$ and $f$ under $\Lambda_1^-$.
Similarly, $p^K$ and $d^{xy}$ transform under $\Lambda_2^-$ 
and $\Lambda_2^+$, respectively.
In the rest of this work, we continue using $\psi_{\pm i}$ and $\eta_{\pm i}$ for time-reversal even bond orders, and introduce primed symbols $\psi'_{\pm i}$ and $\eta'_{\pm i}$ for time-reversal odd current order parameters.
Importantly, even though they contain different form factors,  spatial symmetries including the classification in Table~\ref{tab:C6vtransformations} do not distinguish between $\psi_{\pm i}$ and $\psi'_{\pm i}$, or $\eta_{\pm i}$ and $\eta'_{\pm i}$.
While separating primed and unprimed order parameters is natural in the presence of TRS, the resulting twelve complex numbers are an overparametrization of the problem. 
A natural description can be obtained by setting $\Delta^*_{-i}=\Delta_{+i}\equiv \Delta_i$ and $-\Delta^{\prime *}_{-i}=\Delta^{\prime}_{+i}\equiv \Delta_i'$.
Therefore a single bond (current) along $\pm \vec{Q}_i$ can be parametrized by a complex order parameter $\Delta_i^{(\prime)}=|\Delta_i^{(\prime)}|e^{i\delta^{(\prime)}}$.
Here $\Delta\in\{\psi, \eta\}$ and we denote the corresponding phases by $\delta=\{\phi,\theta\}$.

To gain an intuitive understanding of the various bond and current orders, we consider the Fourier transform of Eq.~(\ref{eq:HMF}) into real space.
Focusing on a single pair of $\pm \vec{Q}_i$ and a single bond or current form factor, this yields
\begin{align}
    H^{f_{\rm b}}_i(\vec{r})=\sum_{\vec{r},\sigma}|\Delta_i| \sum_{\vec{a}} &f_{\rm b}(\vec{a}) (c^\dag_{\vec{r}+\vec{a},\sigma}c_{\vec{r},\sigma} + c^\dag_{\vec{r},\sigma}c_{\vec{r}+\vec{a},\sigma}) \nonumber\\
    &\times\cos[\vec{Q}_i(\vec{r}+\vec{a}/2)+\delta_i],\\
    H^{f_{\rm c}}_i(\vec{r})=\sum_{\vec{r},\sigma}|\Delta'_i| \sum_{\vec{a}} i&f_{\rm c}(\vec{a}) (c^\dag_{\vec{r}+\vec{a},\sigma}c_{\vec{r},\sigma} - c^\dag_{\vec{r},\sigma}c_{\vec{r}+\vec{a},\sigma}) \nonumber\\
    &\times\cos[\vec{Q}_i(\vec{r}+\vec{a}/2)+\delta'_i],
\end{align}
where $f_{\rm b/c}(\vec{a})$ are prefactors depending on both the form-factor and the nearest-neighbor bond $\vec{a}$.
In Figs.~\ref{fig:bonds} and \ref{fig:currents}, we show the real-space pattern of each form factor in the case of a single-$Q$ (only $\pm\vec{Q}_1$) and a multi-$Q$ (equal superposition of all components) ground state. If a single- or multi-$Q$ state is realized depends on the coupling between order parameter components and is determined by higher-order terms in the Landau free energy as we describe below.

Before we move on, let us comment on the analogous theory obtained when $\vec{Q}_i$ fall along the $\Sigma$ line.
In this case, the role of the two sets of mirror operations are exchanged.
Namely, if a representative $\vec{Q}_1\parallel k_x$ [see Fig.~\ref{fig:Lattice}(b)] then the corresponding LG is $\{E,M_y\}$.
More generally, the LG at symmetry-related momenta can be written as $\{E, \sigma_v\}$, with mirror-even and -odd irreps $\Sigma_{1,2}$, which additionally split under TRS as $\Sigma_{1,2}^\pm$.
As the mirror operations exchange roles, now the form factors $p^M$ and $f$ belong to $\Sigma_2^-$ and $p^K$ transforms under $\Sigma_1^-$, while the rest of the classification remains unchanged, as summarized in  Table~\ref{tab:C6vtransformations}.
The corresponding six-component order-parameters transform the same way under space-group symmetries as $\psi_i$ and $\eta_i$, with the caveat that $M_x$ and $M_y$ in Table~\ref{tab:C6vtransformations} are exchanged.
In what follows we construct the Landau free energy in terms of $\psi_i$,  and $\eta_i$, which respectively represent order parameters in $\Lambda_{1}$ or $\Sigma_1$ and $\Lambda_{2}$ or $\Sigma_2$, with the physical form factors following the appropriate classification.

\section{Landau free energy}\label{sec:Landau}

We move on to phenomenologically investigate the above bond and current orders by constructing the corresponding Landau free energies. We start with a general derivation based on symmetries and subsequently connect to a microscopic model as an example realization.
The form of the Landau free energy is valid for hexagonal systems beyond the triangular lattice because it relies on symmetry properties only, which, as we mentioned above, are generalizable with an appropriate redefinition of order parameters and form factors.
We first consider the mirror-even irrep $\Lambda_1^{\pm}$ or $\Sigma_1^\pm$, for which the free energy up to quartic order reads
\begin{align}
    F_{\psi}&=\alpha^\psi \sum_{i=1}^3 |\psi_{i}|^2 + \gamma^\psi \sum_{\tau=\pm} \psi_{\tau 1} \psi_{\tau 2} \psi_{\tau 3}
    + \beta_1^\psi \sum_{i=1}^3|\psi_{i}|^4\nonumber \\
    &+ \beta_2^\psi \Big( |\psi_{1}|^2|\psi_{2}|^2 + |\psi_{2}|^2|\psi_{3}|^2 + |\psi_{1}|^2|\psi_{3}|^2 \Big),\label{eq:L1m} \\
    F_{\psi'}&=\alpha^{\psi'} \sum_{i=1}^3 |\psi'_{i}|^2 + 
     \beta_1^{\psi'} \sum_{i=1}^3|\psi'_{i}|^4\nonumber \\
    &+ \beta_2^{\psi'} \Big( |\psi'_{1}|^2|\psi'_{2}|^2 + |\psi'_{2}|^2|\psi'_{3}|^2 + |\psi'_{1}|^2|\psi'_{3}|^2 \Big),
\end{align}
where $\alpha^{\psi^{(\prime)}}$, $\beta_{1,2}^{\psi^{(\prime)}}$ and $\gamma^\psi$ are phenomenological coefficients. 
The minimum of these free energies determines if the ground state is single- or multi-$Q$. The easiest way to see which configuration minimzes $F_\Delta$ is by rewriting the quartic term as
\begin{align}\label{eq:condition}
    F^{(4)}_\Delta&=\beta_1^\Delta \left( \sum_i \Delta_i^2 \right)^2 \\ 
    & - (2\beta_1^\Delta-\beta_2^\Delta)\Big( |\Delta_{1}|^2|\Delta_{2}|^2 + |\Delta_{2}|^2|\Delta_{3}|^2 + |\Delta_{1}|^2|\Delta_{3}|^2 \Big) \nonumber
\end{align}
with $\Delta\in\{\psi,\psi'\}$.
For $2\beta_1-\beta_2>0$ (where for brevity we suppressed the superindex) the corresponding ground state is an equal superposition of $|\psi^{(\prime)}_{1,2,3}|$, i.e., multi-$Q$.
In contrast, $2\beta_1-\beta_2<0$ favors a unidirectional, (single-$Q$) ground state, breaking rotational symmetry.
Notice that a cubic term is symmetry-allowed for bond orders $\psi_i$.
It is the only term sensitive to the complex phase of the components and can be rewritten as $\gamma^\psi |\psi_1||\psi_2||\psi_3| \cos(\phi_1 + \phi_2 + \phi_3)$.
This term thus locks the sum of the phases to be $0$ ($\pi$) for $\gamma^\psi<0$ ($\gamma^\psi>0$).
However, due to TRS, such term is not allowed in $F_{\psi'}$, where the first phase sensitive term appears at sixth order.

The lowest-order coupling between $\psi_i$ and $\psi_i'$ is of the form
\begin{align}
    F_{\psi \psi'}=\gamma^{\psi \psi'} \Big( 
    &|\psi_1||\psi'_2||\psi'_3| \cos(\phi_1 + \phi'_2 + \phi'_3)\nonumber\\ 
    +&|\psi'_1||\psi_2||\psi'_3| \cos(\phi'_1 + \phi_2 + \phi'_3)\nonumber\\ 
    +&|\psi'_1||\psi'_2||\psi_3| \cos(\phi'_1 + \phi'_2 + \phi_3) \Big).
\end{align}
Being linear in $\psi_i$, this term induces the $\Lambda_1^+$ ($\Sigma_1^+$) bond as secondary order whenever a multi-$Q$ $\Lambda_1^-$ ($\Sigma_1^-$) current is present, and locks the relative phases so as to extremize the cosine functions.
In contrast, if $\psi_i$ condenses at a higher critical temperature than $\psi_i'$, the above term renormalizes 
the quadratic term and introduces a mixing of components $\psi_i'$. 
The renormalized quadratic part then has to be re-diagonalized to identify the instability.
Note that one would generally expect the bond and current sectors to couple via a term linear in perpendicular magnetic field $B$.
However, as $B$ is odd under both $\sigma_d$ and $\sigma_v$, the mirror-even irreps only allow for more trivial coupling terms containing $B^2$, which are insensitive to the field direction.

While we do not commit to a specific momentum along the high-symmetry line, commensurate momenta allow for additional terms that lock the density wave to the lattice.
Let us again consider the $\Lambda$ line as example. 
For a commensurate CDW $|\vec{Q}_{\pm i}|=r |\vec{K}|$ with some rational number $r$.
As an example, if $r=1/2$, then a term $\Delta_{+i}^{(\prime)6}+ \Delta_{-i}^{(\prime) 6}$ is allowed by translational symmetry, as $3 \vec{K}$ falls back onto the $\Gamma$ point.
When expanded, this term contains $|\Delta^{(\prime)}_i|^6 \cos(6 \delta^{(\prime)}_i)$, and quantizes the individual spatial phases $\delta^{(\prime)}_i$ with $\Delta\in\{\psi,\eta\}$, $\delta\in\{\phi,\theta\}$).
Incommensurate momenta, on the other hand, do not allow for such terms, and the overall phase in this case is expected to be pinned by other factors, such as defects.

Next, we examine the mirror-odd sector, transforming under $\Lambda_2$ or $\Sigma_2$.
The free energy containing only $\eta_i$ and $\eta_i'$ up to fourth order is of the form
\begin{align}
    F_{\eta}&=\alpha^\eta \sum_{i=1}^3 |\eta_{i}|^2 + \beta_1^\eta \sum_{i=1}^3 |\eta_{i}|^4 \nonumber \\
    +& \beta_2^\eta \Big(  |\eta_{1}|^2|\eta_{2}|^2 + |\eta_{2}|^2|\eta_{3}|^2 + |\eta_{1}|^2|\eta_{3}|^2  \Big),  \\
    F_{\eta'}&=\alpha^{\eta'} \sum_{i=1}^3 |\eta'_{i}|^2 + \beta_1^{\eta'} \sum_{i=1}^3 |\eta'_{i}|^4 \nonumber \\
    +& \beta_2^{\eta'} \Big(  |\eta'_{1}|^2|\eta'_{2}|^2 + |\eta'_{2}|^2|\eta'_{3}|^2 + |\eta'_{1}|^2|\eta'_{3}|^2  \Big).\label{eq:L2p}
\end{align}
In contrast to $F_{\psi}$, a cubic term in $F_\eta$ is not allowed due to the mirror oddness of $\eta_{\pm i}$, and the first phase sensitive term in  $F_{\eta^{(\prime)}}$ is sixth order.
For the same reason, there is no cubic term coupling $\eta_i$ and $\eta_i'$.
On the other hand, this sector admits a linear  coupling term to the magnetic field $B$, which reads
\begin{align}
    F_{\eta \eta'}= \kappa B \Big(
    &|\eta'_1| |\eta_2| |\eta_3| \sin(\theta'_1+\theta_2+\theta_3) \nonumber \\ 
    +&|\eta_1| |\eta'_2| |\eta_3| \sin(\theta_1+\theta'_2+\theta_3)\nonumber \\
    +& |\eta_1| |\eta_2| |\eta'_3| \sin(\theta_1+\theta_2+\theta'_3) \Big).
\end{align}
Hence, whenever a mirror-odd bond order is present in a multi-$Q$ state, the above term induces a corresponding current as soon as $B\neq0$, simultaneously locking the phases to extremize the sine functions.

Finally, we consider coupling terms between mirror-even and -odd representations.
While a biquadratic coupling is generally symmetry allowed between distinct ordered phases, here we focus on leading-order terms which are linear in at least one of the orders and give rise to secondary or intertwined phases.
As $\psi$ and $\eta$ are differentiated by mirror symmetry, a term quadratic in $\eta$ and linear in $\psi$ is symmetry allowed, so long as the combined momenta fall back on the $\Gamma$ point.
Furthermore, a term containing $\psi \eta'$ or $\psi' \eta$, or vice versa, is allowed, if the mirror and time-reversal oddness is compensated by magnetic field.
Together, the free energy encompassing the coupling of the two representations to cubic order reads
\begin{align}
    &F_{\psi \eta} = \mu \sum_{\langle i,j,k \rangle} |\psi_i||\eta^{(\prime)}_j||\eta^{(\prime)}_k| \cos(\phi_i + \theta^{(\prime)}_j + \theta^{(\prime)}_k)  \\
    & +B \sum_{i=1}^3\Big[ \kappa_{\psi \eta} |\psi'_i| |\eta_i| \sin(\phi'_i-\theta_i) + \kappa_{\eta \psi} |\psi_i| |\eta'_i| \sin(\theta'_i-\phi_i) \Big], \nonumber
\end{align}
where $\langle i,j,k\rangle$ sums over cyclic permutations of $(1,2,3)$.
Again, $\mu$, $\kappa_{\psi \eta}$ and $\kappa_{\eta \psi}$ are coefficients, depending on microscopic details of the system.
Phenomenologically, if $\mu \neq 0$, then a multi-$Q$ bond or current in the mirror-odd sector induces a mirror-even bond as secondary order.
On the other hand, finite $\kappa_{\psi \eta}$ or $\kappa_{\eta \psi}$ results in one bond inducing the opposite current, or vice versa, when a magnetic field is present.

\section{Microscopic model}\label{sec:microscopic_model}

We consider a system described by the extended Hubbard Hamiltonian $\mathcal{H}=\mathcal{H}_0+ \mathcal{H}_U +\mathcal H_J$ on the triangular lattice with
\begin{align}
    \mathcal{H}_0 &= 
    -t 
    \sum_{\braket{i,j}}
    \sum_\sigma 
    \left( 
        c_{i,\sigma}^\dagger 
        c_{j\sigma}^{\phantom{\dagger}} 
        + 
        \text{h.c.} 
    \right) 
    -\mu 
    \sum_i \sum_{\sigma} 
    n_{i,\sigma} 
    \,, \label{eq:H0} \\
    \mathcal{H}_U &= 
    \frac{U}{2} 
    \sum_i\sum_{\sigma,\sigma'} 
    n_{i,\sigma} n_{i,\sigma'} 
    \,, \label{eq:HU} \\
    \mathcal{H}_J &= 
    J 
    \sum_{\langle i,j \rangle} \sum_{\sigma, \sigma'} 
    S^n_{i} S^n_{j} 
    \,, \label{eq:HJ}
\end{align}
where $t$ is the nearest-neighbor hopping amplitude, $\mu$ the chemical potential, $U>0$ represents the onsite repulsive Hubbard interaction, and $J$ the nearest-neighbour exchange interaction. 
We consider a system with a generalized number of fermion 
flavors $\sigma=1,\ldots,N$. 
The density operator for a given flavor $\sigma$ is $n_{i,\sigma}=c_{i,\sigma}^\dagger c_{i\sigma}^{\phantom{\dagger}}$ and the generalised SU(N) spin 
operators are  $S_i^n = c_{i\sigma}^\dagger T^n_{\sigma\sigma'} c_{i\sigma'}^{\phantom{\dagger}}$ with the generators of SU(N) 
$T^n$ in the defining representation and $n=1,\ldots, N^2-1$. 
For SU(2), these are simply proportional to the Pauli matrices $T^n=\sigma^n/2$. 

We investigate CDW instabilities with momenta along the high-symmetry line $\Lambda$ by analyzing divergences of the renormalized interaction within the random-phase approximation (RPA), which is valid for large $N\rightarrow \infty$. This neglects contributions to the charge channel from the crossed particle-hole channel, which are suppressed by $1/N$ \cite{Braun2025,PhysRevB.79.195125,Maiti2013}. 
To set up the RPA, we define the bare particle-hole susceptibility 
\begin{align}
    \Pi_+(\vec{Q}_i) &= 
    -
    \frac{1}{\beta}
    \sum_{i\omega_n}
    \sum_{p}
    \mathcal{G}^0(i\omega_n,\vec{p})
    \mathcal{G}^0(i\omega_n,\vec{p}+\vec{Q}_i),
\end{align}
with $\beta=1/(k_B T)$, Matsubara frequency $\omega_n$, and the non-interacting Green's function $\mathcal{G}^0(i\omega_n,\vec{k})=\bigl(i\omega_n-\epsilon_{\vec{k}}\bigr)^{-1}$.  
We use the short-hand $\int_k\equiv\int_{\mathrm{BZ}}\frac{\mathrm{d}^2\vec{k}}{(2\pi)^2}$ for momentum integrals over the Brillouin zone.  
The tight-binding dispersion on the triangular lattice is
\begin{align}
    \epsilon_{\vec{k}} = -2t\bigl(\cos k_x + 2\cos(k_x/2)\cos(\tfrac{\sqrt{3}}{2}k_y)\bigr) - \mu. \label{eq:dispersion}
\end{align}

Furthermore, we project the interaction and the susceptibility onto an orthonormal set of lattice form factors $\{F_{i,l}(\vec{k})\}$ 
\begin{align}
    \hat{V}^{(l,l')}(\vec{Q}_i) &= 
    \int_{\vec{k}, \vec{k}'} 
    F_{i,l}(\vec{k}+\tfrac{\vec{Q}_i}{2}) 
    F_{i,l'}(\vec{k}'-\tfrac{\vec{Q}_i}{2}) 
    V(\vec{Q}_i;\vec{k},\vec{k}') 
    \,, \label{eq:Vll}\\
    \hat{\Pi}_+^{(l,l')}(\vec{Q}_i) &= 
    -
    \frac{1}{\beta}
    \sum_{i\omega_n}
    \int_{p} 
    \mathcal{G}^0(i\omega_n, \vec{p}) 
    \mathcal{G}^0(i\omega_n, \vec{p}+\vec{Q}_i) \notag \\
    & \times 
    F_{il}\left(\vec{p}+\tfrac{\vec{Q}_i}{2}\right)F_{il'}\left(\vec{p}+\tfrac{\vec{Q}_i}{2}\right)\,,\label{eq:Pill}
\end{align}
where the form factors satisfy the orthonormality and completeness relations $\int_k F_{i,l}(\vec{k}) F_{i,l'}(\vec{k}) = \delta_{l,l'}$ and $\sum_l F_{i,l}(\vec{k}) F_{i,l}(\vec{k}') = \delta(\vec{k}-\vec{k}')$. 
In practice, we choose form factors that transform as the irreducible representations of the triangular-lattice point group $C_{6v}$ and truncate anything beyond nearest neighbors, i.e. $l,l'\in$ \{$s$, $s'$, $f_i$, $p^M_i$, $p^K_i$, $d^{x^2-y^2}_i$, $d^{xy}_i$\}. 
The index $l$ labels the representation while the index $i$ labels the rotation with respect to $Q_i$. To ease notation we suppress the $\pm$ sign, but the vector $Q_i$ can still be any $Q_{\pm1},Q_{\pm2},Q_{\pm3}$ with the appropriate sign choice in $F_{i,l}$, see Sec.~\ref{sec:space_group_symmetry}.

Within RPA the interaction in the CDW channel is then given in matrix form by
\begin{align}
    \hat V_{\mathrm{CDW}}(\vec{Q}_i) = 
    \bigl[
        \mathbf{1} + N\,\hat V(\vec{Q}_i)\hat\Pi_+(\vec{Q}_i)
    \bigr]^{-1}
    \hat V(\vec{Q}_i),
\end{align}
where $\hat V(\vec{Q}_i)$ and $\hat\Pi_+(\vec{Q}_i)$ denote the matrices with elements $\hat V^{(l,m)}(\vec{Q}_i)$, respectively, and $\mathbf{1}$ is the identity matrix. 
An RPA instability (divergence of the CDW ladder) occurs when an eigenvalue of $\;N\,\hat V(\vec{Q}_i)\hat\Pi_+(\vec{Q}_i)\;$ reaches $-1$, from which we can determine the critical temperature $T_c$. 
Because the bare particle--hole bubble is positive semidefinite, this criterion requires a net attractive effective interaction in the corresponding form-factor channel.
In our parametrization this attraction is supplied by the exchange $J$ which leads to attractive components in 
higher angular momentum channels.  
To see this, we perform a Fourier transform and projection of Eqs.~\eqref{eq:HU}-\eqref{eq:HJ} into the nearest-neighbor form factors and obtain
\begin{align}
    \hat V^{(l,l')}(\vec{Q}_i)&=
    \left(
        U 
        -
        \sqrt{3/2}\frac{J}{N}
        s'(\vec{Q}_i)
    \right)
    \delta_{l,s} \delta_{l',s} 
    \nonumber \\
    &\quad
    - 
    \frac{J}{2} 
    \sum_{L \in \mathrm{NN}} 
    \delta_{l,L} \delta_{l',L}\label{eq:projection}\,,
\end{align}
where $\mathrm{NN}$ contains all nearest-neighbor form factors in Eqs.~\eqref{eq:ff1},\eqref{eq:ff2}. We can see that the onsite $U$ feeds into $s$-wave (l=0) channel while the nearest-neighbor exchange $J$ contributes in an attractive way to the nearest-neighbor form-factor channels. 

Note that there are other mechanisms that can produce effective attraction in non-$s$-wave channels even without a bare exchange interaction. One possibility is a spin-fluctuation mechanism \cite{Braun2025}. 
Beyond large-$N$, also other couplings such as a nearest-neighbor interaction can provide an attraction on the bare level via the exchange contribution to the charge channel.

The momentum dependence of the RPA denominator is dominated by the projected particle--hole bubble $\hat \Pi_+(\vec{Q}_i)$. Thus, the CDW instability occurs at the transfer momentum $\vec{Q}_i$ where the leading eigenvalue of $\hat\Pi_+(\vec{Q}_i)$ is maximized as $\hat V(\vec{Q}_i)$ is diagonal in form factors.
At the Van Hove singularity (VHS), this maximum resides at the $M$ points for the s-wave component, as well as for the higher angular momentum components. 
However, doping away from Van Hove filling alters this structure as the single peak near each $M$ point splits and moves toward the $K$ points.
With further doping, peaks originating from neighboring $M$ points merge along the $\Lambda$ high-symmetry line, eventually constituting the global maximum of the susceptibility. 
Consequently, our subsequent analysis focuses on these two distinct regions of parameter space. We provide details of the numerical implementation in App.~\ref{app:RPA}.

\begin{figure}
    \centering
    \includegraphics[width=0.88\columnwidth]{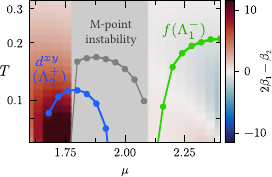}
    \caption{
        Phase diagram 
        as a function of $T$ and $\mu$.
        The blue and green (gray) line show the critical temperature $T_c$ for the CDW instability for $Q_i=K_i/4$ ($Q_i=M_i$) which transform according to the irreps $\Lambda_2^+$ with $d^{xy}$ form factor and $\Lambda_1^-$ with $f$ form factor, respectively.
        The gray area indicates the region where a transition to the $Q_i=M_i$ CDW instability occurs before one towards $Q_i=K_i/4$. 
        The background color encodes $2\beta_1-\beta_2$ for each instability which determines if the order is single-Q ($2\beta_1-\beta_2<0$) or multi-Q ($2\beta_1-\beta_2>0$). 
         The parameters used are $U=3t$, $J=0.7t$ and $N=4$.
    }
    \label{fig:phase_diagram_and_multiQ_criterion}
\end{figure}

Figure \ref{fig:phase_diagram_and_multiQ_criterion} displays the phase diagram as a function of $T$ and $\mu$ obtained from the instability condition that the leading eigenvector of the kernel $\hat V \hat \Pi_+$ equals -1. 
The example parameters used for this phase diagram are $U=3t$, $J=0.7t$ and $N=4$. 
Around the Van Hove singularity, the leading instability is towards a $\vec{Q}_i\approx\vec{M}_i$ CDW as found before \cite{linChiralTwistHigh$T_c$2019,classenCompetingPhasesInteracting2019,Braun2025}.
In agreement with Ref.~\cite{Braun2025}, we find that the leading instability below the Van Hove singularity is towards a $\vec{Q}_i\approx\vec{K}_i/4$ CDW with $d^{xy}_i$-wave form factor at $T_c$.
The corresponding order parameter transforms under the $\Lambda_2^+$ irrep of the LG at $\Lambda$, and represents TRS-preserving mirror-odd bond-density wave order, with the corresponding real-space pattern shown in Fig.~\ref{fig:bonds}.
Above the Van Hove singularity, the leading instability is in the $\Lambda_1^-$ irrep, predominantly in the $f$-wave channel, with the same $\vec{Q}_i=\vec{K}_i/4$ as before.
The ordered phase is therefore mirror-even, TRS-breaking current-density wave, see Fig.~\ref{fig:currents}.

Next, we calculate the coefficients of the Landau free energy from the microscopic model in Eqs.~\eqref{eq:H0}-\eqref{eq:HJ} to determine if the ground state is single- or multi-$Q$. To this end, we decouple the attractive components of the CDW interaction by a Hubbard–Stratonovich field, integrate out fermions, and expand the fermion determinant in the order parameter field $\Delta_l(\vec{Q}_i)$. In the explicit evaluation of the coefficients, we focus on the form factors that correspond to the instability, i.e., $l$ is either $d^{xy}$ or $f$. We provide the details of the evaluation in App.~\ref{app:coefficients} and \ref{app:singlemulti}. 
The quadratic coefficient $\alpha$ is determined by the inverse of the eigenvalue of $\hat V_{\mathrm{CDW}}$ and changes sign at the transition.

The cubic term $\gamma$ of the Ginzburg-Landau free energy 
vanishes for both leading instabilities, as the $d^{xy}$ form factor is odd under mirror symmetry, and the $f$-wave is odd under TRS, such that they forbid a nonzero contraction in the relevant channels.

Finally, the quartic part is given by
\begin{align}
    &\beta_{ijmh} =
    \frac{1}{4\beta}
    \sum_{i\omega_n}
    \int_k
    \mathcal{G}^0(i\omega_n, \vec{k})  \mathcal{G}^0(i\omega_n, \vec{k}+\vec{Q}_i)\label{eq:quartic_matsum} \\
    &\times\mathcal{G}^0(i\omega_n, \vec{k}+\vec{Q}_i + \vec{Q}_j)  \mathcal{G}^0(i\omega_n, \vec{k}+\vec{Q}_i + \vec{Q}_j + \vec{Q}_m) \nonumber\\
   & \times F_{i,l}(\vec{k}+\frac{\vec{Q}_i}{2})
    F_{j,l}(\vec{k}+\vec{Q}_i+\frac{\vec{Q}_j}{2})
    \nonumber \\
    &\times
    F_{m,l}(\vec{k}+\vec{Q}_i+\vec{Q}_j+\frac{\vec{Q}_m}{2})
    F_{h,l}(\vec{k}+\frac{\vec{Q}_i}{2}+\frac{\vec{Q}_j}{2}+\frac{\vec{Q}_m}{2}), \nonumber    
\end{align}
with $\vec{Q}_i+\vec{Q}_j+\vec{Q}_m+\vec{Q}_h=0$. It contains the symmetry-distinct coefficients $\beta_1,\beta_2,\ldots$ that control whether the ordered state is single-$Q$ or multi-$Q$, see Eqs.~\eqref{eq:L1m}--\eqref{eq:L2p}. 
The different combinations of $i,j,m,h$ can be grouped into two classes, which result in the two different $\beta$ coefficients $\beta_1$ and $\beta_2$.
In particular the sign of the combination $2\beta_1-\beta_2$ determines the preference: if $2\beta_1-\beta_2>0$ multi-$Q$ order is favored, whereas $2\beta_1-\beta_2<0$ favors single-$Q$ order, see Eq.~\eqref{eq:condition}.
The numerical evaluation of these coefficients for the leading eigenvectors 
is shown in Fig.~\ref{fig:phase_diagram_and_multiQ_criterion}.
Below the Van Hove singularity \LC{$\mu\lesssim 1.75t$} the quantity $2\beta_1-\beta_2$ is positive for the temperatures considered, therefore multi-$Q$ order is favored. 
Combined with the symmetry of the leading eigenvector this yields a multi-$Q$ $d^{xy}_i$ bond-density wave in that regime.
Above the Van Hove singularity $\mu\gtrsim 2.1t$ the quantity $2\beta_1-\beta_2$ near $T_c$ changes sign as a function of $\mu$, indicating a change in the ground state from multi$-Q$ to unidirectional $f$-wave current-density wave for increasing $\mu$. 
Note that the multi-$Q$ phase in the low-temperature regime around $2.1t\leq \mu \leq 2.2 t$ may not be stable and higher orders need to be taken into account in the Landau free energy  (see App.~\ref{app:singlemulti}), potentially inducing a first-order transition.

\section{Conclusion}
We argued that CDW can be comprised of non-trivial local textures of bonds or currents in addition to a modulation with wave vector $\vec{Q}$ and its star, i.e., its symmetry-related wave vectors. We showed that these local textures can be classified by irreps of the little group at $\vec{Q}$ and developed a framework that simultaneously takes the symmetries of the little group and the multi-component nature of a CDW induced by the star into account. 

As an application case, we considered two-dimensional hexagonal lattices and ordering wave vectors along the $\Lambda$ ($\Gamma-K$) and $\Sigma$ ($\Gamma-M$) lines.
For concreteness, we considered a triangular lattice as a minimal example. Incorporating time-reversal symmetry, we demonstrated that there are four classes of order parameters which transform according to the two mirror-even or -odd irreps $\Lambda_1^\pm$ and $\Lambda_2^\pm$ ($\Sigma_1^\pm$ and $\Sigma_2^\pm$) of the little group at a point on the $\Lambda$ ($\Sigma$) line being even (+) or odd (-) under TRS.
The time-reversal even orders describe bond-density waves, while the time-reversal odd ones describe loop-current phases.

Our symmetry classification allowed us to derive the Landau free energy for all four cases. We also considered the coupling between bond and current orders within the same irrep, as well as the coupling of orders in different irreps. From the free energies, we can determine if a CDW is single- or multi-Q depending on quartic coefficients. The free energies also revealed that a $\Lambda_1^+$ bond order is typically induced by a multi-Q order from mirror-even currents in $\Lambda_1^-$ or mirror-odd bond or currents in $\Lambda_2^\pm$. Furthermore, we discussed the behavior in a magnetic field which enables further couplings between bond and current orders.

As a microscopic realization we considered a triangular-lattice Hubbard model with SU(N) exchange interaction. We performed an RPA instability analysis and derived the coefficients of the Landau free energy for the orders indicated by the instability to determine if they are single- or multi-Q. We found that bond or current orders with non-trivial local textures arise when the chemical potential is in the vicinity of but not at the Van Hove energy. For a chemical potential below the Van Hove energy, we find a multi-Q bond order described by a $d^{xy}$ modulation falling into $\Lambda_2^+$, and for chemical potential above the Van Hove energy we obtain a single-Q current order with local $f$-wave modulation transforming according to $\Lambda_1^-$.

While we used the simple triangular lattice as a concrete example, our symmetry analysis and the corresponding free energies are valid for more complex lattices with hexagonal symmetry, such as the honeycomb or kagome lattices.
Furthermore, related CDW orders have been reported in transition-metal dichalcogenides such as $2H$-TaSe$_2$~\cite{Moncton1977}. 

Our symmetry classification highlights the importance of local textures in complex CDW materials and can be straightforwardly extended to other space groups. CDWs are typically detected by extra Bragg peaks due to translation symmetry breaking, which, however, cannot distinguish between on-site charge and non-local bond or current modulation. Bond and current orders can be distinguished by breaking of time-reversal symmetry with the caveat that we showed that they can occur simultaneously. However, in principle, imaging techniques can also resolve local modulations, and it will be interesting to work out precise detection methods for the non-trivial CDWs discussed in this work in future.

\section*{Acknowledgments}

We thank Dennis Huang for discussions. AS and LC were supported by a grant from the Simons Foundation SFI-MPS-NFS-00006741-11.

\bibliography{bibliography}

\appendix 

\section{Numerical evaluation of the RPA susceptibility and Ginzburg-Landau coefficients}
\label{app:numerics}
 
This appendix summarizes the numerical procedure underlying the RPA eigenvalue/eigenvector analysis and the Ginzburg-Landau (GL) coefficients quoted in the main text.

\subsection{Lattice, dispersion, and Brillouin-zone integration}
 
The triangular-lattice tight-binding dispersion is implemented as
\begin{align}
    \epsilon_k = -2t\left[\cos k_x + 2\cos\tfrac{k_x}{2}\cos\tfrac{\sqrt3 k_y}{2}\right] - \mu \,,
\end{align}
with $t=1$ setting the energy scale, consistent with $\mathcal H_0$ in the main text. 
Momenta are parametrized in fractional reciprocal-lattice coordinates, $\vec{k} = x\,  \vec{b}_1 + y\, \vec{b}_2$ with $x,y\in[0,1]$ and
\begin{align}
    \vec{b}_1 = \left(2\pi, -\tfrac{2\pi}{\sqrt3}\right), \qquad
    \vec{b}_2 = \left(0,\ \tfrac{4\pi}{\sqrt3}\right),
\end{align}
so that all Brillouin-zone (BZ) integrals reduce to integrals of $x,y$ over the unit square, i.e. over the primitive-cell parallelogram, which has the same area and periodicity as the hexagonal BZ. 
All quadrature routines are evaluated with tight tolerances (relative tolerance $10^{-8}$--$10^{-9}$, absolute tolerance $10^{-10}$--$10^{-12}$, and evaluation budgets of up to $10^{9}$ integrand evaluations), using adaptive cubature for the full parameter sweeps described below.

\subsection{RPA susceptibility, interaction matrix, and leading instability} \label{app:RPA}
 
For a given momentum transfer $\vec{Q}_i$, temperature $T$, and chemical potential $\mu$, the bare particle-hole bubble is projected onto the form-factor basis as a $7\times7$ matrix because $l,l'\in$ \{$s$, $s'$, $f_i$, $p^M_i$, $p^K_i$, $d^{x^2-y^2}_i$, $d^{xy}_i$\},
\begin{align}
    \hat\Pi_+^{(l,l')}(\vec{Q}) &= 
    -
    \int_{k} 
    F_{il}\left(k+\tfrac{\vec{Q}}{2}\right) 
    F_{il'}\left(k+\tfrac{\vec{Q}}{2}\right)
    \nonumber \\
    &\quad
    \times
    \frac{n_F(\epsilon_{\vec{k}+\vec{Q}})-n_F(\epsilon_{\vec{k}})}{\epsilon_{\vec{k}+\vec{Q}}-\epsilon_{\vec{k}}} \,,
    \label{eq:bubble_analytic}
\end{align}
i.e., the Lindhard bubble is evaluated in closed form (the Matsubara sum over the two internal fermionic frequencies is carried out analytically), leaving only the momentum integral to be performed numerically.
To avoid the numerically ill-conditioned $0/0$ limit as $\epsilon_{k+Q}\to\epsilon_{k}$, the difference quotient is replaced, whenever $|\epsilon_{k+Q}-\epsilon_{k}|$ falls below a threshold of $10^{-5}$, by its Taylor expansion in $\delta\epsilon\equiv\epsilon_{k+Q}-\epsilon_{k}$ up to second order,
\begin{align}
    \frac{n_F(\epsilon+\delta\epsilon)-n_F(\epsilon)}{\delta\epsilon} \approx n_F'(\epsilon) + \tfrac12 n_F''(\epsilon)\,\delta\epsilon + \tfrac16 n_F'''(\epsilon)\,\delta\epsilon^2 \,,
\end{align}
with the derivatives of the Fermi function evaluated analytically from $n_F(\epsilon)=[e^{\epsilon/T}+1]^{-1}$.
 
The corresponding interaction matrix $\hat V(\vec{Q})$ is diagonal in the form-factor basis in the present implementation, i.e. every channel receives the exchange contribution $-J/2$, while the momentum-independent ($s$-wave) channel additionally carries the bare repulsion $U>J>0$ together with a correction $-\frac{\sqrt3}{2}\frac{J}{N}s'(\vec{Q})$ evaluated at the ordering momentum itself, see Eq.~(\ref{eq:projection}) of the main text. 
 
The RPA kernel is formed as $\hat\kappa(\vec{Q}) = \hat V(\vec{Q})\hat\Pi_+(\vec{Q})$ and diagonalized directly to obtain its eigenvalues $\lambda_n(\vec{Q})$ and eigenvectors $\vec v_n(\vec{Q}) \in \mathbb R^7$. 
The channel with the most negative eigenvalue is identified as the leading instability; its eigenvector gives the linear combination of the seven basis form factors, $f(\vec{k})=\sum_a v_{n,a}\, f_a(\vec{k})$, that defines the order-parameter form factor entering the GL free energy.
There is only little mixture of different form factors in the leading eigenvector, so the resulting order parameter is dominated by a single lattice irrep (e.g. $d^{xy}_i$ or $f_i$) as discussed in the main text.
As the interaction matrix is diagonal in the present implementation, the eigenvectors of $\hat\kappa(\vec{Q})$ are close to identical to those of $\hat\Pi_+(\vec{Q})$, and the eigenvalues differ only by a channel-dependent prefactor given by the corresponding diagonal element of $\hat V(\vec{Q})$.

 \subsection{Evaluation of coefficients in the Landau free energy}
 \label{app:coefficients}

We compute the quadratic, cubic and quartic coefficients appearing in the Landau free energy by evaluating diagrams 
multiplying $n=2,3,4$ order-parameter fields. We obtain the diagrammatic expressions through a Hubbard-Stratonovich decoupling of the action corresponding to Eqs.~\eqref{eq:H0}-\eqref{eq:HJ}, and subsequently integrating out the fermions and expanding in the order-parameter fields.
Each field carries one of  $\{ Q_{\pm 1},Q_{\pm 2},Q_{\pm 3} \}$ momenta, with the requirement that they sum to zero, $\sum_{r=1}^n Q_{i_r}=0$. 
These momenta are obtained 
by a systematic 
search over all $6^n$ 
combinations filtering for the zero-sum condition. 
Because many 
of the allowed momentum combinations are related by the residual $C_6$ rotation symmetry of the label set and by cyclic permutation of the loop (which leaves the enclosed diagram invariant), the resulting list is subsequently grouped into equivalence classes: for each zero-sum 
set, all six rotated copies and all cyclic permutations thereof are generated, and a canonical representative is chosen for each resulting orbit. 

Each equivalence class is recorded together with its size (i.e., its multiplicity in the original enumeration), so that only one representative integral per class needs to be evaluated numerically, weighted by the class size when the full sum is reassembled.
For the quadratic coefficient, there is only one class of two-propagator loops, $(Q_i,Q_{-i})$, with multiplicity $6$. The explicit expression is given by
\begin{align}
    \alpha_{l,l'}(\vec{Q}_i) &= 
    \frac{1}{2}[\hat V^{-1}(\vec{Q}_i) + N \hat \Pi_+(\vec{Q}_i)]_{l,l'} \,,
    \label{eq:quadratic_matsum}
\end{align}
with $\hat V$ and $\hat\Pi_+$ defined in the main text Eqs.\eqref{eq:Vll}, \eqref{eq:Pill}. The quadratic coefficient $\alpha_{l,l'}(\vec{Q}_i)=\alpha_{l,l'}$ is equal for all $Q_i$ and thus does not favor any specific direction of the momentum transfer. For the Hubbard and exchange interaction we consider, $\hat V_{l,l'}=\mathrm{diag}(\lambda_l)$ is diagonal with $\lambda_s=\left(U -\sqrt{3/2}\frac{J}{N}s'(\vec{Q}_i)\right)$ and $\lambda_l=-J/2$ for $l\in$\{$s$, $s'$, $f_i$, $p^M_i$, $p^K_i$, $d^{x^2-y^2}_i$, $d^{xy}_i$\}. Thus, $\alpha_{l,l'}=(\delta_{l,l'}+N\lambda_l [\hat\Pi_+(\vec{Q}_i)]_{l,l'})/\lambda_l$, which means that a phase transition occurs when an eigenvalue of $N\lambda_l [\hat\Pi_+(\vec{Q}_i)]_{l,l'}$ reaches $-1$. Note that this criterion is equivalent to the instability analysis within RPA. In the following evaluation of the cubic and quartic coefficients, we consider only the ones that correspond to the order parameter field of this instability, i.e., we do not explicitly evaluate the coupling terms between different orders.

The cubic coefficients are given by
\begin{align}
    \gamma_{ijh} &= 
    -\frac{1}{3\beta}
    \sum_{i\omega_n}
    \int_k
    \mathcal{G}^0(i\omega_n, k)
    F_{i,l}(k+Q_i/2)
    \label{eq:cubic_matsum}\\
    &\quad \times
    \mathcal{G}^0(i\omega_n, k+Q_i)
    F_{j,l}(k+Q_i+Q_j/2)
    \nonumber \\
    &\quad \times
    \mathcal{G}^0(i\omega_n, k+Q_i+Q_j)
    F_{h,l}(k+Q_i/2+Q_j/2) \,, \nonumber    
\end{align}
where $l$ denotes the form factor corresponding to the instability. Finally, the expression for the quartic coefficients is given in the main text Eq.~\eqref{eq:quartic_matsum}.

For the cubic and quartic GL coefficients, which involve products of three and four single-particle Green's functions along a closed momentum loop, the fermionic Matsubara sum
\begin{align}
    T\sum_{i\omega_m} \prod_{r=1}^{n} \frac{1}{i\omega_m-\epsilon_r}
\end{align}
is evaluated by direct numerical summation over $2N_{\rm mat}=2000$ Matsubara frequencies $\omega_m = (2m+1)\pi T$, $m=-N_{\rm mat},\dots,N_{\rm mat}-1$, rather than via the closed-form residue expression.

For the cubic coefficient, there are two classes of three-propagator loops, represented by the two "mercedes-star" combinations $(Q_1,Q_2,Q_3)$ and $(Q_1,Q_3,Q_2)$ (see Fig.~\ref{fig:Lattice}), each with multiplicity $6$.
For the quartic coefficient, there are five classes of four-propagator loops, represented by the two self-contractions $(Q_1,Q_{-1},Q_1,Q_{-1})$ and $(Q_1,Q_1,Q_{-1},Q_{-1})$ with multiplicities $6$ and $12$, respectively, and the four distinct mixing contributions 
$(Q_1,Q_{-3},Q_{-1},Q_3)$, 
$(Q_1,Q_3,Q_{-3},Q_{-1})$, 
$(Q_1,Q_{-3},Q_{3},Q_{-1})$ and 
$(Q_1,Q_{2},Q_{-2},Q_{-1})$ with multiplicities $12$, $12$, $24$ and $24$, respectively.
 
\subsection{Single- vs. multi-Q order}
 \label{app:singlemulti}
For each RPA eigenchannel identified, the quadratic, cubic, and quartic GL coefficients are evaluated as BZ integrals of the corresponding closed fermionic loop as described in the previous section. 
The quartic coefficients are then grouped into the two classes (self and mixed contractions), which are denoted $\beta_1$ and $\beta_2$ in the main text.
\begin{align}
    \beta_1 &= \beta_{1,-1,1,-1} + 2\,\beta_{1,1,-1,-1}\,, \\
    \beta_2 &= \beta_{1,-3,-1,3} + \beta_{1,3,-3,-1} \nonumber\\
    &\quad+ 2\,\beta_{1,-3,3,-1} + 2\,\beta_{1,2,-2,-1}\,,
\end{align}
where $\beta_{i_1,\dots,i_4}$ denotes the four-propagator loop integral built from the momentum labels $(Q_{i_1},\dots,Q_{i_4})$, see Eq.~\ref{eq:quartic_matsum}.
A final factor of $6$ is multiplied to both $\beta_1$ and $\beta_2$.

The criterion for stability of the ordered state is that the quartic part of the GL free energy is positive definite, which requires $\beta_1>0$ and $2\beta_1 + \beta_2>0$.
\begin{figure}[h]
    \centering
    \includegraphics[width=0.5\linewidth]{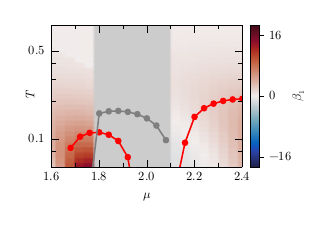}
    \caption{Quartic Ginzburg-Landau coefficient $\beta_1$ for $Q=K/4$, evaluated for the leading RPA eigenchannel as a function of temperature and chemical potential.
    The red (grey) line indicates the critical temperature $T_c$ for the CDW instability for $Q=K/4$ ($Q=M$).
    }
    \label{fig:app_beta1}
\end{figure}

\begin{figure}[h]
    \centering
    \includegraphics[width=0.5\linewidth]{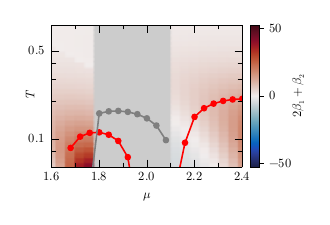}
    \caption{Phase-stability criterion $2\beta_1+\beta_2$ for $Q=K/4$, evaluated for the leading RPA eigenchannel as a function of temperature and chemical potential.
    The red (grey) line indicates the critical temperature $T_c$ for the CDW instability for $Q=K/4$ ($Q=M$).}
    \label{fig:app_stability}
\end{figure}
The criterion for multi-Q vs single-Q order is
\begin{align}
    2\beta_1-\beta_2\, > 0 \quad \text{(multi-Q)}, \qquad
    2\beta_1-\beta_2\, < 0 \quad \text{(single-Q)}\,.
\end{align}
All integrals sharing the same list of momentum labels are cached and evaluated only once per parameter point $(T,\mu)$. 
The full coefficient set is computed on a grid of $21$ chemical-potential values linearly spaced over $\mu\in\{1.6,2.4\}\,t$ (bracketing $\mu_{\rm VH}=2.0\,t$) and $49$ temperatures log-spaced over $T\in\{10^{-2},1\}\,t$, with $U=3.0\,t$ and $J=0.7\,t$ for $N=4$ fermion flavors, using multi-threaded parallelization over the parameter grid. We report $2\beta_1-\beta_2$ for the two considered instabilities at $Q=K/4$ in Fig.~\ref{fig:phase_diagram_and_multiQ_criterion} of the main text. In Figs.~\ref{fig:app_beta1} and \ref{fig:app_stability} we also show the evaluation of $\beta_1$ and $2\beta_1+\beta_2$, which are positive in the considered parameter range, i.e., the considered phases are stable.

\end{document}